\documentclass{article}

\usepackage[utf8]{inputenc}
\usepackage[T1]{fontenc}
\usepackage{amsmath, amssymb}
\usepackage[colorlinks=false, linkcolor=black, citecolor=black, urlcolor=black, hidelinks]{hyperref}
\usepackage{geometry}
\usepackage{graphicx}
\usepackage{authblk}
\usepackage[numbers,square]{natbib}
\usepackage{placeins}
\usepackage{algorithm, algorithmicx}
\usepackage{algpseudocode}

\title{\textbf{A coupled fully kinetic hydrogen transport and ductile phase-field fracture framework for modeling hydrogen embrittlement}\\
{\small Published in: \href{https://doi.org/10.1016/j.ijplas.2026.104765}{International Journal of Plasticity 204 (2026) 104765}}}

\author[1]{Abdelrahman Hussein\thanks{Corresponding author: abdelrahman.hussein@oulu.fi; ORCID=0000-0001-9458-6484}}
\author[2]{Yann Charles}
\author[1]{Jukka Kömi}
\author[1]{Vahid Javaheri}
\affil[1]{Materials and Mechanical Engineering, Faculty of Technology, Center for Advanced Steels Research, University of Oulu, Pentti Kaiteran katu 1, Oulu, 90570, Finland}
\affil[2]{Université Sorbonne Paris Nord, Laboratoire des Sciences des Procédés et des Matériaux, LSPM, CNRS, UPR 3407, F-93430, Villetaneuse, France}
\date{} 

\begin{document}

\maketitle

\begin{abstract}
  Modeling hydrogen embrittlement (HE) is a long-standing engineering challenge, which has experienced significant developments in recent years. Yet, there is a gap in modeling the effect of the kinetics of hydrogen segregation at dislocations and the resulting interaction between ductile tearing and hydrogen-induced brittle fracture. In this work, a comprehensive chemo-mechanical framework is developed by coupling the fully kinetic hydrogen transport model with the geometric phase-field fracture method. A novel driving force is proposed that utilizes a hyperbolic tangent function of stress triaxiality to ensure that plastic dissipation contributes to fracture only under tensile conditions, phenomenologically representing void-driven ductile damage. The model successfully predicts the hydrogen-dependent shift in damage initiation from the specimen core to the surface. More importantly, hydrogen segregation at dislocations was shown to be crucial for modeling the multiple surface cracking experimentally observed at the necking region. Furthermore, the framework captures the competition between loading rates and diffusion kinetics, resolving the transition from multiple circumferential surface cracking at high strain rates to center-initiated single crack at lower rates. Finally, the model reproduced the experimental J-resistance curves for compact tension specimens, showing the transition from ductile tearing to embrittled crack.
\end{abstract}


\section{Introduction}

Hydrogen Embrittlement (HE) is an environmentally assisted damage mechanism in metallic materials, where the infusion of solute hydrogen triggers catastrophic failure through a premature loss of fracture toughness at loading levels significantly below nominal design limits. While several mechanisms have been proposed in the literature for HE, notably Hydrogen Enhance Decohesion (HEDE) and Hydrogen Enhanced Localized Plasticity (HELP) \cite{singh2022coupled, park2025modeling, tondro2022effects, taherijam2025hydrogen, tekkaya2025coupled}, they follow a broadly similar sequence of chemo-mechanical interactions from a continuum perspective: hydrogen ingress into the material, followed by diffusion and accumulation at tensile stressed regions and microstructure—such as dislocations and interfaces—which ultimately degrades their local load-carrying capacity below nominal levels. Therefore, a robust numerical framework must accurately resolve the coupling between these three fundamental aspects of HE, which ultimately govern fracture.

Stress-driven diffusion is often modeled by assuming Oriani's local equilibrium between the lattice and dislocation-trapped hydrogen populations \cite{Sofronis1989, Krom1999, DiLeo2013}. In this framework, while the total concentration is partitioned between these sites, only lattice hydrogen is assumed mobile. Consequently, the effective transport kinetics are governed by the diffusion rate of the mobile lattice hydrogen. On the other side, the McNabb-Foster formulation has been employed to capture the transient exchange kinetics between the trapped and diffusive populations \cite{diaz2019numerical, charles2021effect, park2024continuum}; however, hydrogen at dislocations is still treated as immobile. Recent studies calculated the flux of the trapped species by the mobility of the trapping sites in the case of dislocations \cite{dadfarnia2015modeling, charles2022modeling} and vacancies \cite{chroeun2025modeling}. The stationary assumption inherent in classical trapping models precludes the representation of phenomena such as pipe diffusion, where microstructural defects act as high-speed transport pathways. To account for the effects of dislocation mobility on hydrogen transport, a drift flux term was added using dislocations velocity vector \cite{dadfarnia2015modeling} and gradient of plastic strain using strain gradient plasticity \cite{yuan2022key, lindblom2025strain}. Recently, a fully kinetic formulation has been recently developed that treats hydrogen accumulation at microstructural defects through a diffusion-based mechanism rather than traditional reaction kinetics. This framework has been applied for segregation at grain boundaries \cite{Hussein2024,Hussein2024b}, thermal desorption spectroscopy in duplex stainless steel \cite{Hussein2025TDS} and dislocations \cite{Hussein2025DIS}. The latter incorporates the spatial gradient of the normalized dislocation density directly into the driving force for transport, the model captures the segregation of hydrogen as a continuous flux toward regions of high solubility, effectively enabling modeling phenomena like pipe diffusion.

The locally brittle aspects of HE have been generally modeled using Cohesive Zone Models (CZM) \cite{serebrinsky2004quantum, olden2009influence, yu2016viscous}. Conversely, ductile damage in metals—predominantly driven by void nucleation, growth, and coalescence—is widely represented by Gurson-Tvergaard-Needleman (GTN) models \cite{shang2023unraveling, Besson2020crack} within the framework of continuum damage mechanics (CDM). This approach has been implemented to investigate the ductile aspects of HE in several studies \cite{yu2019hydrogen, Pinto2024simulation, lin2024experimental, fernandez2026oxygen}, while the ductile-to-brittle transition has been modeled through a combination of CZM and GTN formulations \cite{lin2022simulation}. However, CZM necessitates a predetermined crack path, and GTN models require complex non-local gradient-based formulations to mitigate mesh sensitivity \cite{tuhami2022two}. Furthermore, the aforementioned models typically utilize a hydrogen trapping formulation based on Oriani's local equilibrium at dislocations, thereby limiting the hydrogen accumulation rate to the lattice diffusion and neglect transient kinetics within plastically deforming regions—crucial for the observed loading rate effects on HE \cite{krom1999effect}.

The phase-field fracture method is a gradient-based approach that does not require a predetermined crack path \cite{Miehe2010_IJNME, de2016gradient}. Because it is founded on an energetic crack driving force—traditionally based on the elastic strain energy density $\psi$—and incorporates an inherent length scale $\ell$ for regularization, it has gained significant interest for modeling the complex interactions in HE \cite{martinez2018phase, isfandbod2021mechanism, singh2024hydrogen}. However, standard formulations often struggle to capture the influence of plastic deformation, which has been partially mitigated by incorporating a fraction (e.g., 10\%) of the plastic work density into the crack driving force \cite{mandal2025computational}.

Furthermore, current phase-field formulations typically neglect the explicit hydrogen accumulation at dislocations, relying instead on the apparent diffusivity to account for trapping effects. This approach fails to resolve the localized increase in hydrogen solubility at dislocation sites—a microstructural feature widely recognized in the literature as critical for the initiation and propagation of HE damage \cite{lin2024experimental, kim2024pre, lee2025direct}. Rather, it was assumed that sub-critical cracks are responsible for the increased hydrogen concentration within the sample as shown by recent modeling efforts \cite{martinez2020suitability, cui2024computational}. However, such behavior is largely a consequence of the model's numerical formulation rather than a reflection of the underlying chemo-mechanics. In these frameworks, the phase-field order parameter $\phi$ initiates damage ($\phi>0$) almost immediately upon loading at stress concentrators. This initiation, in conjunction with the penalty boundary conditions imposed at the crack front, creates a moving boundary that effectively transports environmental hydrogen into the bulk via the advancing crack tip. While this methodology can be calibrated for notched geometries, it is less suited to capture HE characteristics in specimens with uniform cross-sections, such as smooth tensile bars, where fracture initiation is governed dislocation evolution and transient local accumulation hydrogen rather than geometric effects.

Ductile damage has been formulated within the phase-field fracture framework by Ambati et al. \cite{Ambati2015ductile, Ambati2016phase} by replacing the standard quadratic degradation function of the AT2 regularization with a function coupled to the plastic state represented by the equivalent plastic strain normalized by a critical threshold $\varepsilon_\mathrm{eq}/\varepsilon_\mathrm{eq,crit}$. This formulation produces two significant effects: first, it delays crack initiation until significant plastic strain accumulates; and second, it ensures that crack propagation is driven by plastic strain rather than only the elastic strain energy. While this approach successfully captures characteristic ductile patterns, the resulting evolution equation is non-linear in the phase-field variable $\phi$ and is not extensible to account for other effects often required for ductile damage like stress triaxiality. 

Miehe et al. \cite{Miehe2015_p1} presented a variational geometric phase-field formulation where the crack evolution is governed by a modular crack driving state function. This framework allows for a generic driving force to be utilized without altering the underlying finite element formulation and directly integrates with an operator split \cite{Miehe2010_CMAME} for a robust staggered linearized implementation. For ductile materials, the damage driving force is typically derived from plastic work density or accumulated plastic strain. Crucially, this approach is modular in its treatment of damage initiation; it can be implemented in a no-threshold sense, similar to AT2 regularization \cite{kristensen2021assessment}, or in a threshold sense by defining an explicit critical value that must be overcome before phase-field evolution begins. A mechanistic-based GTN based formulation was later used for porous plasticity using void evolution for crack driving force \cite{miehe2016phase, Alakheel2018}. Borden et al. \cite{borden2016phase} proposed a phenomenological ductile fracture model that integrates triaxiality and Lode angle effects through a state-dependent plastic strain threshold. By utilizing a standard $J_2$ plasticity framework, this approach avoids the numerical instabilities associated with the non-isochoric yield surfaces of GTN-like models. Instead, it maintains computational robustness by shifting the stress-state dependency into a failure envelope that triggers phase-field evolution only after a critical, triaxiality-dependent plastic strain is reached.

In this work, we present a chemo-mechanical framework developed to resolve the kinetic interactions between solute hydrogen, dislocation evolution, and material degradation. Central to this approach is the fully-kinetic transport formulation, which utilizes the normalized dislocation density as a thermodynamic driving force for the transient accumulation of hydrogen \cite{Hussein2025DIS}. The dislocation density is calculated using a Kocks-Mecking-Estrin \cite{Estrin1984} evolution model, which could be seamlessly integrated with various plastic hardening laws to accommodate a wide range of metallic systems and deformation behaviors. For ductile damage, we build upon the modular geometric phase-field framework by proposing a simplified ductile driving force that implicitly captures the stress-state dependency of porous metals without the numerical complexities of non-isochoric yield surfaces. By scaling the plastic dissipation with a triaxiality-dependent hyperbolic tangent function, the formulation ensures that the plastic work density contributes to phase-field evolution only under favorable tensile conditions, mimicking the physics of void growth. While not exhaustive GTN-like models in representing the mechanistic details of void growth and coalescence, it successfully captures the essential features of ductile damage using a minimal set of parameters, allowing for a significantly more efficient numerical implementation. Indeed, this framework is able to capture characteristic ductile damage features like cup-and-cone morphology in notched samples. Furthermore, when extended to hydrogen-embrittlement studies, the model effectively characterizes the transition to surface cracking in the central regions of smooth round bar tensile samples and accurately predicts the associated loss of fracture toughness in compact tension (CT) geometries.

\section{Governing equations}

\subsection{Geometric phase-field fracture}
\label{Section: PFF}

The geometric approach to phase-field fracture proposed by Miehe et al. \cite{Miehe2015_p1} is an elegant formulation that allows flexibility and modularity in constructing crack initiation and propagation driving force without altering the underlying finite element weak form, and thus, the discretization procedures. Using the phase-field order parameter, which represents a smooth transition from undamaged ($\phi=0$) to a fully damaged region ($\phi=1$), the regularized crack surface density functional is expressed as

\begin{equation}
    \Gamma_\ell(\phi) = \int_V \, \Gamma(\phi, \nabla \phi) dV, \qquad \text{with} \qquad \Gamma(\phi, \nabla \phi) = \frac{1}{2\ell} (\phi^2 + \ell^2 |\nabla \phi|^2).
    \label{Eq: CrackSurfDen}
\end{equation}

\noindent where $\ell$ is a length-scale parameter determining the smearing of the crack. The variational derivative of Eq.~(\ref{Eq: CrackSurfDen}) reads

\begin{equation}
    \delta_\phi \Gamma_\ell(\phi) = \frac{1}{\ell} (\phi - \ell^2 \Delta \phi) = \frac{1}{\ell} \mathcal{D}_\mathrm{c} \quad \text{in} V,
    \label{Eq: VarDeriv}
\end{equation}

\noindent with the natural boundary condition on the surface $\partial V$

\begin{equation}
    \nabla \phi \cdot \mathbf{n} = 0 \quad \text{on } \partial V.
\end{equation}

\noindent where $\mathcal{D}_\mathrm{c}$ is the dimensionless \emph{geometric} crack resistance defined as

\begin{equation}
    \mathcal{D}_\mathrm{c} = \ell \, \delta_\phi \Gamma_\ell = \phi - \ell^2 \Delta \phi
\end{equation}

\noindent The essence of the geometric approach is describing crack evolution by the balance of a \emph{crack driving force} and the geometric resistance. Ignoring viscous terms, this can be expressed as \cite{Miehe2015_p1}

\begin{equation}
    \phi - \ell^2 \Delta \phi = (1 - \phi) \, \mathcal{H}
    \label{Eq: CrackBalance}
\end{equation}

\noindent where $\mathcal{H}$ is the crack driving force, with the \emph{state} dependence

\begin{equation}
    \mathcal{H} = \max_{s \in [0, \, t]} \tilde{\mathcal{D}}(state, s)
\end{equation}

\noindent In what follows, variables with tilde accent ($\sim$), such as $\tilde{\mathcal{D}}$, strictly represent undegraded quantities or expressions derived from them. To enforce irreversibility, the evolution of $\mathcal{H}$ is governed by the Kuhn--Tucker conditions

\begin{equation}
    \dot{\mathcal{H}} \ge 0 \, , \quad
    \mathcal{H} \ge \tilde{\mathcal{D}} \, , \quad
    \dot{\mathcal{H}} (\mathcal{H} - \tilde{\mathcal{D}}) = 0 \, .
\end{equation}

\noindent The evolution of the phase-field variable $\phi$ is then governed by the complementary system

\begin{equation}
    \dot{\phi} \ge 0 \, , \quad (1 - \phi)\, \mathcal{H} - \mathcal{D}_\mathrm{c}   \le 0 \, , \quad \dot{\phi} \left[(1 - \phi)\, \mathcal{H} - \mathcal{D}_\mathrm{c}\right] = 0 \, .
\end{equation}

\noindent Eq.~(\ref{Eq: CrackBalance}) shows that a crack will evolve only when the \emph{effective} crack driving force equals the geometric crack resistance, i.e., when $(1 - \phi)\, \mathcal{H} = \mathcal{D}_\mathrm{c}$, in accordance with the Kuhn--Tucker conditions. The flexibility of Eq.(\ref{Eq: CrackBalance}) in constructing a crack driving force $\tilde{\mathcal{D}}$, compared to the energetic Griffith-like formulation, which allows only the elastic work density as the crack driving force, becomes evident.

In their original work \cite{Miehe2015_p1, Miehe2015_p2}, Miehe et al. used the traditional Griffith-like crack driving force $\tilde{\mathcal{D}}^\mathrm{e}$ based on the elastic strain energy density in addition to proposing two elastoplastic crack driving forces with a threshold $\langle \tilde{\mathcal{D}}^\mathrm{ep} \rangle$ and without a threshold $\tilde{\mathcal{D}}^\mathrm{ep}$, defined as 

\begin{equation}
    \tilde{\mathcal{D}}^\mathrm{e} = \frac{\tilde{\psi}^\mathrm{e+}}{w_\mathrm{c}} \, , \qquad
    \langle \tilde{\mathcal{D}}^\mathrm{ep} \rangle = \zeta \left\langle \frac{\tilde{\psi}^\mathrm{e+} + \tilde{w}^\mathrm{p}}{w_\mathrm{c}} - 1 \right\rangle \, , \qquad
    \tilde{\mathcal{D}}^\mathrm{ep} = \frac{(\tilde{\psi}^\mathrm{e+} + \tilde{w}^\mathrm{p})}{w_\mathrm{c}} \, .
\end{equation}

\noindent where $\langle x \rangle := \max(x, 0)$ is the Macaulay bracket, $\zeta$ is a parameter that controls the post critical behavior, $w_\mathrm{c}$ is the critical work density and is related to the critical energy release rate via $w_\mathrm{c} = G_\mathrm{c}/2\ell$, $\tilde{\psi}^\mathrm{e+}$ is the positive part of the strain energy density, ensuring that damage evolves only in tensile state. The split of the strain energy density can be expressed as \cite{Miehe2010_CMAME}

\begin{equation}
    \tilde{\psi}^{\mathrm{e}+}
    = \frac{1}{2}\lambda \langle \mathrm{tr}(\boldsymbol{\varepsilon}^\mathrm{e}) \rangle^2_+ + \mu \mathrm{tr}([\boldsymbol{\varepsilon}^\mathrm{e+}]^2) \quad  \text{and} \quad 
    \tilde{\psi}^{\mathrm{e}-}
    = \frac{1}{2}\lambda \langle \mathrm{tr}(\boldsymbol{\varepsilon}^\mathrm{e}) \rangle^2_- + \mu \mathrm{tr}([\boldsymbol{\varepsilon}^\mathrm{e-}]^2) 
    \label{Eq: ElasticDecomp}
\end{equation}

\noindent where $\lambda$ and $\mu$ are the Lamé parameters, with $\langle x \rangle_+ := \max(x, 0)$ and $\langle x \rangle_- := \min(x, 0)$. The split of the elastic strain tensor is based on the spectral decomposition

\begin{equation}
    \boldsymbol{\varepsilon}^{\mathrm{e}+}
    = \sum_{i=1}^{n} \langle \varepsilon^\mathrm{e}_i \rangle_+ \, \mathbf{n}_i \otimes \mathbf{n}_i \, \, ,
    \qquad \text{and} \qquad
    \boldsymbol{\varepsilon}^{\mathrm{e}-}
    = \sum_{i=1}^{n} \langle \varepsilon^\mathrm{e}_i \rangle_- \, \mathbf{n}_i \otimes \mathbf{n}_i \, .
\end{equation}

\noindent where $\varepsilon^\mathrm{e}_i$ are the principal strains and $\mathbf{n}_i$ are principal directions. $\tilde{w}^\mathrm{p}$ is the plastic work density, defined as 

\begin{equation}
    \tilde{w}^\mathrm{p} = \int_0^t \boldsymbol{\tilde{\sigma}} \!:\! \dot{\boldsymbol{\varepsilon}}^\mathrm{p} \, ds
\end{equation}

\noindent where $\boldsymbol{\tilde{\sigma}}$ is the \emph{undegraded} stress and $\dot{\boldsymbol{\varepsilon}}^\mathrm{p}$ is the plastic strain rate. Observe that $\tilde{w}^\mathrm{p}$ is derived from undegraded quantities, ensuring that it is a monotonically increasing function, reflecting the accumulation of plastic work. Finally, the embrittlement effect of hydrogen can be introduced by a decrease in the critical work density $w_\mathrm{c}$ 

\begin{equation}
  w_\mathrm{c}(\bar{c}) = w_\mathrm{cmin} + (w_\mathrm{c0} - w_\mathrm{cmin}) \exp( -\beta \bar{c})
  \label{Eq: wc}
\end{equation}

\noindent where $\bar{c}$ is the hydrogen concentration normalized by a critical value $c_\mathrm{crit}$, $w_\mathrm{c0}$ is the critical work density without hydrogen, i.e. when $c = 0$. $w_\mathrm{cmin}$ is the maximum embrittlement due to hydrogen. Adjusting this value can determine whether the damage evolves in the elastic or elastoplastic range, thereby controlling the degree of ductile to brittle transition. $\beta$ is a parameter to control the steepness of the embrittlement from $w_\mathrm{c0}$ to $w_\mathrm{cmin}$.

\subsection{A ductile criterion for crack driving force}
\label{Section: DuctilePFF}

As discussed in Eq.~(\ref{Eq: ElasticDecomp}), $\tilde{\psi}^{\mathrm{e}+}$ accounts for the tensile contribution to the driving force. Specifically, the association with the positive principal strain term $\langle \mathrm{tr}(\boldsymbol{\varepsilon}^\mathrm{e}) \rangle_+$, which captures the tensile volumetric expansion, implicitly resembles the role of stress triaxiality in GTN-like void-driven ductile damage models. On the other hand, $\tilde{w}^\mathrm{p}$ provides the contribution of the plastic dissipation to damage evolution that characterizes ductile failure. Therefore, for a better phenomenological representation of ductile damage, yet maintaining model simplicity, we propose the following ductile damage driving force

\begin{equation}
    \langle \tilde{\mathcal{D}}^\mathrm{d} \rangle = \zeta \left\langle \frac{\eta \, \tilde{\psi}^\mathrm{e+} + (1-\eta) \, \tanh (\kappa \, \langle \mathcal{T}  \rangle) \, \tilde{w}^\mathrm{p}}{w_\mathrm{c}} - 1 \right\rangle
    \label{Eq: Ductile}
\end{equation}

\noindent where $\eta$ is a weighing factor that determines the contributing ratio $\tilde{\psi}^{\mathrm{e}+}/\tilde{w}^\mathrm{p}$, $\mathcal{T}$ is the stress triaxiality defined by the ratio of the hydrostatic stress to the von Mises equivalent stress $= \sigma_\mathrm{h}/\sigma_\mathrm{eq}$, and $\kappa$ is a prefactor that determines the sensitivity to triaxiality. The term $\tanh \, (\kappa \, \langle \mathcal{T}  \rangle)$ allows $\tilde{w}^\mathrm{p}$ to contribute to $\langle \tilde{\mathcal{D}}^\mathrm{d} \rangle$ only in a state of positive triaxiality. We use $\kappa=6$ such that $\tanh (\kappa \, \langle \mathcal{T}  \rangle) \approx 0.964$ for a state of uniaxial tension, i.e. $\mathcal{T} = 1/3$.

Therefore, the proposed formulation Eq.~(\ref{Eq: Ductile}) implicitly captures the underlying physics of void-driven ductile damage in metals without having to implement the complex, non-isochoric yield surface of the GTN-like models that are challenging for numerical convergence \cite{miehe2016phase, Alakheel2018}. In these models, the coupling of pressure-dependent plastic dilatation with material softening severely degrades the conditioning of the consistent tangent operator, requiring artificial viscous regularization to preserve numerical stability \cite{Pinto2024simulation}. By contrast, the present formulation decouples the hydrostatic stress dependency from the constitutive plastic update, where the physical manifestation of void growth under tensile triaxiality is instead implicitly evaluated through the modular tensile crack driving force via the spectral decomposition of the elastic strain energy. Because the plastic flow rule remains strictly isochoric, complying with the standard $J_2$ framework, the resulting local return-mapping equations remain well-conditioned.

\subsection{Elastoplasticity}
\label{Section: Elastoplasticity}
The degradation accompanied by damage evolution is described by 

\begin{equation}
  g(\phi)_\mathrm{d} = (1-\phi)^2
\end{equation}

\noindent The evolution of stress follows the conservation of linear momentum 

\begin{equation}
  \text{div} \, \boldsymbol{\sigma} = \mathbf{0}, \qquad \boldsymbol{\sigma} = g(\phi)_\mathrm{d} \, \tilde{\boldsymbol{\sigma}} = g(\phi)_\mathrm{d} \, \mathbf{C}^\mathrm{e} \!:\! \boldsymbol{\varepsilon}^\mathrm{e} \, .
\end{equation}

\noindent A standard small-strain associative, rate-independent $J_2$ plasticity framework is used, wherein the yield function $f$ and the plastic multiplier $\lambda$ (with $\dot{\lambda} = \dot{\varepsilon}_{\mathrm{eq}}$ being the equivalent plastic strain rate) are updated via an implicit backward Euler return-mapping scheme for each iteration $(k)$ according to

\begin{equation}
    \begin{split}
    \Delta \lambda &= \frac{f^{(k)}}{3G \, g(\phi)_\mathrm{d} + H(\lambda^{(k)})}  \\ 
    \lambda^{(k+1)} &= \lambda^{(k)} + \Delta \lambda^{(k+1)} \\
    f^{(k+1)} &= \tilde{\sigma}_\mathrm{eq}^\mathrm{tr} - 3G \, g(\phi)_\mathrm{d} \, \Delta \lambda - \sigma_\mathrm{y}^{(k+1)}
    \end{split}
    \label{Eq: RMPFF}
\end{equation}

\noindent where $\tilde{\sigma}_\mathrm{eq}^\mathrm{tr}$ is the undegraded trial stress, $G$ is the shear modulus, and $H(\lambda) = d\sigma_{\mathrm{y}}/d\lambda$ is the plastic hardening modulus. In this study, two isotropic hardening laws were utilized to describe the material behavior: Power-law (Swift) hardening and Voce exponential hardening. The Power-law hardening is expressed as

\begin{equation}
    \sigma_{\mathrm{y}} = K \left( \varepsilon_0 + \varepsilon_{\mathrm{eq}} \right)^n, \quad 
    H(\lambda) = \frac{d\sigma_{\mathrm{y}}}{d\varepsilon_{\mathrm{eq}}} = n K \left( \varepsilon_0 + \varepsilon_{\mathrm{eq}} \right)^{n-1}
    \label{Eq: PowerLawHard}
\end{equation}

\noindent where $K$ is the material strain hardening coefficient, $n$ is the strain hardening exponent, and $\varepsilon_0$ is the initial strain offset parameter related to the initial yield strength of the material ($\sigma_{\mathrm{y}0} = K\varepsilon_0^n$). Alternatively, the Voce hardening law and its corresponding hardening modulus are defined as

\begin{equation}
    \sigma_{\mathrm{y}} = \sigma_{\mathrm{sat}} + (\sigma_{\mathrm{y}0} - \sigma_{\mathrm{sat}}) \exp \left( -\frac{H_0}{\sigma_{\mathrm{sat}} - \sigma_{\mathrm{y}0}} \varepsilon_{\mathrm{eq}} \right),
    \quad
    H(\lambda) = \frac{d\sigma_{\mathrm{y}}}{d\varepsilon_{\mathrm{eq}}} = H_0 \exp \left( -\frac{H_0}{\sigma_{\mathrm{sat}} - \sigma_{\mathrm{y}0}} \varepsilon_{\mathrm{eq}} \right)
    \label{Eq: VoceHardening}
\end{equation}

\noindent where $\sigma_{\mathrm{y}0}$ is the initial yield strength, $\sigma_{\mathrm{sat}}$ is the saturation stress, and $H_0$ is the initial strain hardening modulus. Finally, the continuum tangent can be expressed as

\begin{equation}
    \mathbf{C}_\mathrm{ep} = \Big( \mathbf{C}_\mathrm{e} - \frac{\mathbf{C}_\mathrm{e} : \boldsymbol{N}^\mathrm{tr} \otimes \boldsymbol{N}^\mathrm{tr} : \mathbf{C}_\mathrm{e}}{\frac{2}{3} H + \boldsymbol{N}^\mathrm{tr} : \mathbf{C}_\mathrm{e} : \boldsymbol{N}^\mathrm{tr}} \Big) \, g(\phi)_\mathrm{d}
    \label{Eq: IsoHardTangStiffness}
\end{equation}

\noindent where $\boldsymbol{N}^\mathrm{tr} = \frac{\partial f}{\partial \boldsymbol{\sigma}}$ is the plastic flow direction.

It should be stressed that the current framework utilizes a small strain formulation that couples the hydrogen concentration field exclusively to the fracture-resistance threshold $w_{\mathrm{c}}$ without altering the underlying macro-scale plastic flow rule or hardening parameters. Consequently, the simulations do not capture hydrogen-induced plastic softening (such as HELP effects), rate-dependent viscoplastic yield, or post-necking geometric softening behavior. While this introduces some simplifications, it serves as a robust framework to evaluate ductile phase-field fracture for HE.

\subsection{Hydrogen transport}
\label{Section: Trnasport}

The derivation of the fully kinetic hydrogen mass transport equation, including the effects of hydrostatic stress and dislocations, is detailed in \cite{Hussein2025DIS}. Here, we only summarize the main equations and highlight the modifications introduced to account for phase-field damage evolution. The hydrogen flux equation is expressed as

\begin{equation}
  \mathbf{J} = -  \mathbf{D} \left(\nabla c - \frac{\bar{V}_\mathrm{H} \, c}{RT} \nabla \sigma_\mathrm{h} - \frac{\zeta_\rho \, c}{RT} \nabla \bar{\rho} \right)
  \label{Eq: ConFlux}
\end{equation}

\noindent where $\mathbf{D}$ is the diffusivity tensor, $c$ is the hydrogen concentration, $R$ is the universal gas constant, $T$ is the absolute temperature, $\sigma_\mathrm{h}$ is the hydrostatic stress. $\zeta_\rho$ is the hydrogen-dislocation segregation enthalpy, which is analytically related to the occupancy-based segregation enthalpy $\Delta H_{\rho}$ as discussed in detail in \cite{Hussein2025DIS}. The selected value of $\zeta_{\rho} = 13.8\text{ kJ/mol}$ in Table \ref{Table: SimParams}, Section \ref{Section: Staggered} maps to an occupancy-based segregation enthalpy of $\Delta H_{\rho} = -20\text{ kJ/mol}$ \cite{Hussein2025DIS}, which is well within the physically accepted range for ferritic steels. $\bar{\rho}$ is the normalized dislocation density. The dislocation density evolution is calculated from the Kocks-Mecking-Estrin \cite{Estrin1984} dislocation evolution equation, with the closed form solution \cite{Hussein2025DIS}

\begin{equation}
  \rho(\varepsilon_\mathrm{eq}) = \left[ \frac{k_1}{k_2} - \left( \frac{k_1}{k_2} - \sqrt{\rho_0} \right) \exp\Big({-\frac{M k_2\varepsilon_\mathrm{eq}}{2}} \Big) \right]^2
  \label{Eq: KMEAnalytical}
\end{equation}

\noindent where $\varepsilon_\mathrm{eq} = \sqrt{\frac{2}{3}\varepsilon^{p}_{ij}:\varepsilon^{p}_{ij}}$ is the equivalent plastic strain, $\rho_0$ is the initial dislocation density, $k_1$ and $k_2$ are the dislocation multiplication and annihilation coefficients respectively, and $M$ is the Taylor factor for polycrystals. The dislocation density is normalized by the saturation value $\rho_s = (k_1/k_2)^2$, i.e. $\bar{\rho} = \rho/\rho_s$. Assuming isotropic diffusivity, $D = D^x = D^y = D^z$, the diffusivity field can be expressed as  

\begin{equation}
  D = D_\mathrm{L}\Big(1 + m\bar{\rho}\Big)\Big(1 - 0.99 \, \phi^2\Big) \, , \, m \in \, ]-1, \infty[
  \label{Eq: Diffusivity}
\end{equation}

\noindent where $D_\mathrm{L}$ is the diffusivity in a perfect lattice. The parameter $m$ governs the enhancement or retardation of hydrogen along dislocation cells--a homogenized continuum approximation of the collective dislocation population as they physically aggregate into cellular substructures under plastic deformation. Furthermore, in this study, $m$ was set to zero to isolate the kinetic influence of $\nabla \bar{\rho}$ driving force without introducing additional unconstrained diffusivity parameters. The second term, $\left(1 - 0.99  \phi^2\right)$, accounts for a reduction in diffusivity within the damaged regions, where $\phi = 1$ corresponds to a fully degraded material. It should be noted that Eq. (\ref{Eq: Diffusivity}) utilizes an effective, homogenized isotropic diffusivity field. While this approach captures the magnitude variations of transport induced by microstructural segregation and damage degradation, it serves as a macro-scale continuum approximation that does not explicitly resolve the directional anisotropy of transport along individual dislocation cores. Explicitly modeling such anisotropic effects requires resolving discrete dislocation line orientations and their corresponding Burgers vectors, which remains outside the scope of the present $J_2$ continuum framework and is restricted to discrete dislocation or atomistic scale frameworks. The time evolution of concentration can be obtained from the mass conservation as 

\begin{equation}
    \frac{\partial c}{\partial t} - \nabla \cdot \mathbf{D} \, \Big(\nabla c - \frac{\bar{V}_\mathrm{H} \, c}{RT} \nabla \sigma_\mathrm{h} - \frac{\zeta_\rho \, c}{RT} \nabla \bar{\rho} \Big) + Z_d \, \phi^2 \, \Big(c - c_\mathrm{eq}\Big) = 0
    \label{Eq: MassCon}
\end{equation}

\noindent where the last term $Z_d \, \phi^2 \, (c - c_\mathrm{eq})$ is a source/sink term that models the exchange of hydrogen at damaged regions ($\phi > 0$) with the surrounding environment. The choice of quadratic term via $\phi^2$ is to provide efficient localization of the sink term by making it sharper at the crack surface, while it rapidly decays in the bulk undamaged material compared to a linear dependence using only $\phi$. This form models a first-order reaction rate, with $Z_d$ being the rate constant. The term drives the local hydrogen concentration $c$ toward an equilibrium boundary value $c_\mathrm{eq}$ \cite{Hussein2025DIS} defined as 

\begin{equation}
    c_\mathrm{eq} = c_B \exp\Big( \frac{\bar{V}_\mathrm{H} \, \sigma_\mathrm{h} + \zeta_\rho \, \bar{\rho}}{RT} \Big)
    \label{Eq: Conb}
\end{equation}

\noindent where $c_B$ is a constant boundary concentration of the hydrogen environment. When $c_\mathrm{eq} = 0$, the term acts as a pure sink, simulating hydrogen desorption at the crack surface. This condition is applied for pre-charged samples tested in hydrogen-free environment, where hydrogen desorption occurs at the crack surface. On the other hand, when $c_\mathrm{eq} > 0$, the term acts as a source, modeling hydrogen absorption/desorption from a hydrogen-rich environment into the crack surface.

\section{Finite element discretization}
\label{Section: FEM}

\subsection{Phase-field fracture}
\label{Section: PFFFEM}

Multiplying Eq.~(\ref{Eq: CrackBalance}) by a test function $\eta_\phi$, the weak form can be expressed as

\begin{equation}
  \int_V \phi \, \eta_\phi + \ell^2 \nabla \phi \cdot \nabla \eta_\phi \, dV - \int_V (1 - \phi) \mathcal{H} \, \eta_\phi \, dV = 0
\end{equation}

\noindent Using the vector of the shape functions $\mathbf{N}$ and their spatial derivative matrix $\mathbf{B} =\mathbf{\nabla}\mathbf{N}$, the integration point values of $\phi$, $\nabla \phi$ are expressed using their corresponding vector of nodal values as

\begin{equation}      
  \phi = \mathbf{N \,\phi} \, ,  \qquad 
  \nabla \phi = \mathbf{B} \, \mathbf{\phi} \, . 
\end{equation}

\noindent Eliminating the arbitrary $\eta_\phi$ and substituting Eq.~(\ref{Eq: DerivMat}), the residual of the phase-field fracture

\begin{equation}
    \mathbf{R}_\phi =  \int_V (\mathbf{N}\boldsymbol{\phi} \, \mathbf{N}^\top  + \ell^2 \, \mathbf{B}^\top \mathbf{B} \, \boldsymbol{\phi}) dV - \int_V \mathcal{H} (1-\mathbf{N}\boldsymbol{\phi}) \, \mathbf{N}^\top dV
    \label{Eq: PhiResidual}
\end{equation}

\noindent By adopting a semi-implicit time integration scheme of $\mathcal{H}$, i.e., updating only in the mechanical problem while maintained constant in the phase-field problem, Eq.~(\ref{Eq: PhiResidual}) becomes linear in $\phi$. This is an advantage of the geometric phase-field fracture approach \cite{Miehe2015_p1, Miehe2015_p2}. The linear system of equations for phase-field fracture then reads

\begin{equation}
  \begin{split}
    &\int_V \Big[ (\mathcal{H}+1)\mathbf{N}^\top \mathbf{N}  + \ell^2 \, \mathbf{B}^\top \mathbf{B} \Big] \, dV \boldsymbol{\phi} - \int_V \mathcal{H} \, \mathbf{N}^\top \, dV = 0\\
    &\mathbf{K}_{\phi} \boldsymbol{\phi} - \mathbf{f}_\mathcal{H} = 0 \\
    \label{Eq: PFFDiscrete}
  \end{split}
\end{equation}

\subsection{Hydrogen transport}
\label{Section: TransportFEM}

The weak form of Eq.~(\ref{Eq: MassCon}) is obtained by multiplying with a test function $\eta_c$, applying the divergence theorem to the second term and integrating over the domain $V$

\begin{equation}
  \begin{split}
  \int_V \eta_c \, \frac{\partial c}{\partial t} \, dV 
  + \int_V \nabla \eta_c \cdot \mathbf{D} \left( 
      \nabla c - \frac{\bar{V}_\mathrm{H} \, c}{RT} \nabla \sigma_\mathrm{h} 
      - \frac{\zeta_\rho \, c}{RT} \nabla \bar{\rho} 
  \right) \, dV 
  &+ \int_V Z_d \, \phi^2 \, \eta_c \, c \, dV 
  - \int_{A} \eta_c \, \mathbf{J} \cdot \mathbf{n} \, dA \\
  &- \int_V Z_d \, \phi^2 \, \eta_c \, c_\mathrm{eq} \, dV = 0
  \end{split}
  \label{Eq: WeakForm}
\end{equation}

\noindent It should be noted that $\boldsymbol{\sigma}_\mathrm{h}$ and $\bar{\boldsymbol{\rho}}$ are the nodal values mapped from the integration points of the mechanical sub-problem as discussed in \cite{Hussein2025DIS}. In this procedure, the integration-point values of $\sigma_\mathrm{h}$ and $\bar{\rho}$ are mapped onto the mesh nodes via a volume-weighted nodal projection scheme. Using the vector of the shape functions $\mathbf{N}$ and strain matrix $\mathbf{B}$, the integration point values of $c$, $\nabla c$, $\nabla \sigma_\mathrm{h}$ and $\nabla \bar{\rho}$ are expressed using their corresponding vector of nodal values as

\begin{equation}      
  c = \mathbf{N \,c} \, ,  \qquad 
  \nabla c = \mathbf{B} \, \mathbf{c} \, , \qquad 
  \nabla \sigma_\mathrm{h} = \mathbf{B} \, \boldsymbol{\sigma}_\mathrm{h} \, , \qquad
  \nabla \bar{\rho} = \mathbf{B} \, \bar{\boldsymbol{\rho}} \, .
  \label{Eq: DerivMat}
\end{equation}

\noindent Eliminating the arbitrary $\eta_c$ and substituting in Eq.~(\ref{Eq: WeakForm})
  
\begin{equation}
  \begin{aligned}
    \int_V \mathbf{N}^\top\mathbf{N} \frac{\partial\mathbf{c}}{\partial t} \, dV 
    + \int_V \mathbf{B}^\top \mathbf{D} \mathbf{B} \, \mathbf{c} \, dV \,
    &- \int_V \mathbf{B}^\top \Big( \mathbf{D} \frac{\bar{V}_\mathrm{H}}{RT} \Big) \mathbf{B} \, \boldsymbol{\sigma}_\mathrm{h} \, \mathbf{N} \, \mathbf{c} \, dV \,-\int_V \mathbf{B}^\top \Big( \mathbf{D} \frac{\zeta_\rho}{RT} \Big) \mathbf{B} \, \bar{\boldsymbol{\rho}} \, \mathbf{N} \, \mathbf{c} \, dV  \\
    & +\int_V \mathbf{N}^\top\mathbf{N} Z_d \, \phi^2 \mathbf{c} \, dV = \int_{A} \mathbf{N}^\top \mathbf{J} \cdot \mathbf{n} \, dA \, +\int_V \mathbf{N}^\top\mathbf{N} Z_d \, \phi^2 \mathbf{c}_\mathrm{eq} \, dV
  \end{aligned}
  \label{Eq: Discret}
\end{equation}

\noindent The mass $\mathbf{M}$, diffusivity $\mathbf{K_\mathrm{D}}$, interaction $\mathbf{K_\mathrm{T}}$ and sink $\mathbf{K_\mathrm{S}}$ matrices can be expressed as 

\begin{equation}
  \begin{split}
    &\mathbf{M} = \int_V \mathbf{N}^\top \mathbf{N} \, dV \, , \qquad
    \mathbf{K_\mathrm{D}} = \int_V \mathbf{B}^\top \mathbf{D} \mathbf{B} \, dV \, , \\
    &\mathbf{K_\mathrm{T}} = \int_V \mathbf{B}^\top \Big(\mathbf{D}\frac{\bar{V}_\mathrm{H}}{RT}\Big) \mathbf{B} \, \boldsymbol{\sigma}_\mathrm{h} \mathbf{N} \, dV + \int_V \mathbf{B}^\top \Big(\mathbf{D}\frac{\zeta_\rho}{RT} \Big)\mathbf{B} \, \bar{\boldsymbol{\rho}} \, \mathbf{N} \,dV \, , \\
    & \mathbf{K_\mathrm{S}} = \int_V \mathbf{N}^\top\mathbf{N} Z_d \, \phi^2 \, dV \, .
  \end{split}
\end{equation}

\noindent And the nodal RHS load vector 

\begin{equation}
    \mathbf{F} = \int_{A} \mathbf{N}^\top \mathbf{J} \cdot \mathbf{n} \, dA  \, +\int_V \mathbf{N}^\top\mathbf{N} \frac{Z_d}{t} \, \phi^2 \mathbf{c}_\mathrm{eq} \, dV
\end{equation}

\noindent Substituting in Eq. (\ref{Eq: Discret})

\begin{equation}
    \mathbf{M} \, \dot{\mathbf{c}} + (\mathbf{K_\mathrm{D}} - \mathbf{K_\mathrm{T}} + \mathbf{K_\mathrm{S}}) \, \mathbf{c} = \mathbf{F}
    \label{Eq: TimeDisc}
\end{equation}

\noindent Using the implicit Euler time integration to Eq. (\ref{Eq: TimeDisc}), the FE linear system of equations becomes

\begin{equation}
    [\mathbf{M} + \Delta t \, (\mathbf{K_\mathrm{D}} - \, \mathbf{K_\mathrm{T}} + \mathbf{K_\mathrm{S}})] \, \mathbf{c}^{t+1} = \Delta t \, \mathbf{F} + \mathbf{M} \, \mathbf{c}^t
    \label{Eq: TrapTransient}
\end{equation}

\subsection{Computational setup}
\label{Section: Staggered}

\begin{figure}[hbt!]
  \includegraphics[width=10cm]{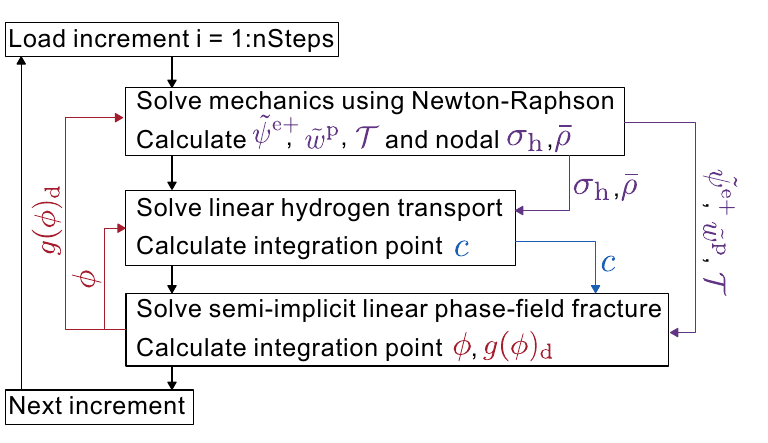}
  \centering
  \caption{Schematic of the staggered solution scheme implemented in PHIMATS for the chemo-mechanical problem. Arrows indicate the variables for inter-physics coupling.}
  \label{Fig: Algorithm}
\end{figure}

A staggered coupling scheme is adopted for the present chemo-mechanical framework, as illustrated in Fig.~(\ref{Fig: Algorithm}). The strategy employs an operator-splitting approach \cite{Miehe2010_CMAME} where the non-linear mechanical sub-problem is solved first, followed by the hydrogen transport, and finally the phase-field fracture within each load increment. 

To ensure computational efficiency, inter-physics coupling variables at the integration points are stored as C++ pointers for high-performance exchange between sub-problems, while nodal-point variables are exchanged by accessing values stored in the output files. This coupled framework is implemented in the open-source code PHIMATS \cite{PHIMATS} and is hosted on GitHub https://github.com/ahcomat/PHIMATS. Selected simulation scenarios discussed in Section~\ref{Section: RD} are provided in the CaseStudies/HydrogenDuctilePFF directory. For all simulations, the actively damaging regions are meshed with mesh size $h = \ell/5$, where $\ell = 3 \times 10^{-2}$ mm in all simulations. All models were meshed with 4-node quadrilateral elements either with axisymmetric or plane strain formulation. Additionally, the material parameters shown in Table \ref{Table: SimParams} are used for all simulations. 

\begin{table}[hbt!]
\caption{Material parameters used in the simulations}
\begin{tabular}{rlrl}

  \hline

  \multicolumn{2}{c}{Mechanical parameters} & 
  \multicolumn{2}{c}{Hydrogen transport parameters} \\

  \hline

  Young's modulus $E$ & 210 GPa & hydrogen-dislocation segregation parameter $ \zeta_\rho$ & 13.805 $\mathrm{kJ/mol}$  \\
  Poisson's ratio $\nu$ & 0.3 &  Partial molar volume of hydrogen $\bar{V}_\mathrm{h}$ & 2 $\times 10^{-6} \, \mathrm{m^3}$ \\
  Dislocation multiplication coefficient $k_1$ & 133$\times 10^6$ & Universal gas constant $R$ & 8.31 $\mathrm{J/molK}$ \\
  Dislocation annihilation coefficient $k_2$ & 10 & Temperature T & 300 $\mathrm{K}$ \\
  Taylor factor $M$ & 3 & Rate constant $Z_d$ & 1e-2 $\mathrm{s^{-1}}$   \\
  Initial dislocation density $\rho_0$ & 1 $\times 10^{11} \, \mathrm{m^{-2}}$ & &  \\

  \hline

\end{tabular}
\label{Table: SimParams}
\end{table}

\section{Results and discussion}
\label{Section: RD}

\subsection{Representative examples for ductile damage}
\label{Section: DuctileDamage}

\begin{figure}[hbt!]
  \includegraphics[width=15cm]{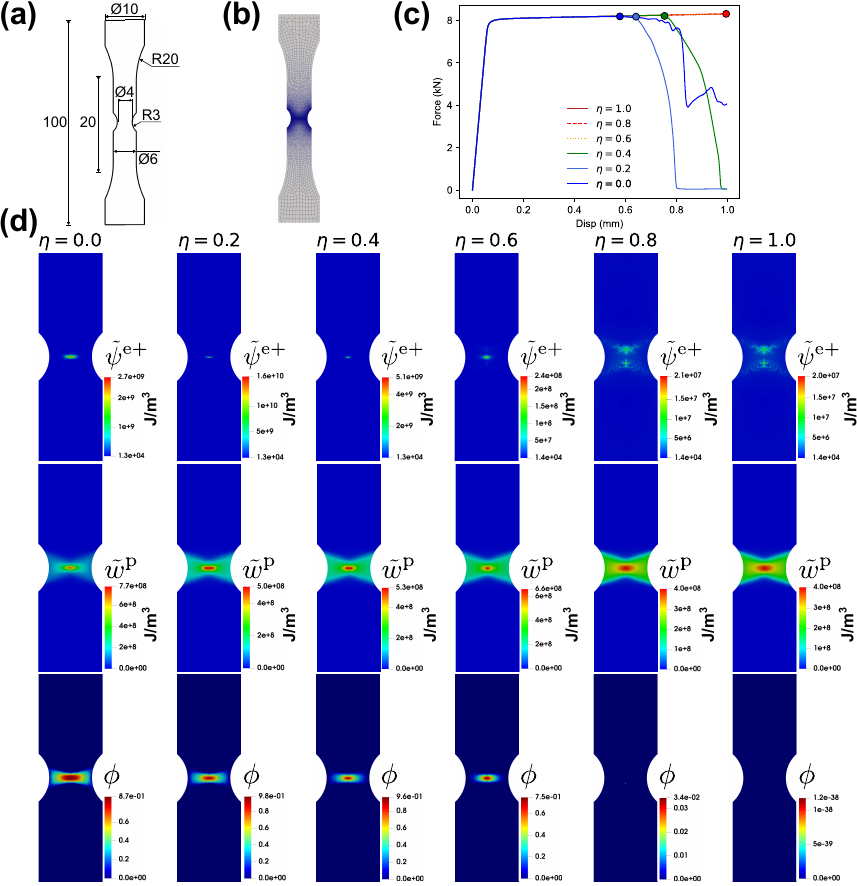}
  \centering
  \caption{(a) Geometry and dimensions of the notched round bar specimen. (b) Finite element mesh. (c) Influence of the weighing factor $\eta$ on the force-displacement curves and (d) plots of $\tilde{\psi}^{\mathrm{e}+}$, $\tilde{w}^\mathrm{p}$ and $\phi$ fields at the loading points highlighted by the markers in (c).}
  \label{Fig: Sensitivity}
\end{figure}

While the study is primarily concerned with the coupled chemo-mechanical modeling of hydrogen embrittlement, this section demonstrates the capability of the proposed driving force in Eq.~\eqref{Eq: Ductile} to capture ductile damage features in the absence of hydrogen. The first verification case considers a Notched Round Bar (NRB) specimen, with the geometry adapted from Ambati et al. \cite{Ambati2016phase} and illustrated in Fig.~(\ref{Fig: Sensitivity}a) together with the finite element mesh in Fig.~(\ref{Fig: Sensitivity}b). A power-law hardening model was used with initial yield strength $\sigma_{\mathrm{y}0} = 500$ MPa, strain hardening coefficient $K=300$ MPa, and hardening exponent $n=0.3$. The critical work density was set to $w_\mathrm{c} = 100 \, \mathrm{MJ/m^3}$.

First, we investigate the effect of the weighting factor $\eta$ on the mechanics of damage initiation and the resulting macro-response of the NRB specimen, as shown in Fig.~(\ref{Fig: Sensitivity}c) and (\ref{Fig: Sensitivity}d). Setting $\eta=0.0$ completely eliminates the contribution of the tensile elastic strain energy density $\tilde{\psi}^{\mathrm{e}+}$ to the crack driving force, leaving the failure process driven entirely by the plastic dissipation density $\tilde{w}^\mathrm{p}$. Damage localizes and initiates at the center of the specimen where the plastic work accumulation is maximum. It should be noted that the extremely high magnitudes (in the $\text{GJ/m}^3$ range) observed for $\tilde{\psi}^{\mathrm{e}+}$ are a direct consequence of the undegraded formulation where $\tilde{\psi}^{\mathrm{e}+}$ is evaluated using the undamaged elastic properties.

Furthermore, the force-displacement response for $\eta=0.0$ exhibits numerical oscillations during the post-peak load drop. This instability highlights that incorporating a portion of the tensile elastic energy acts as an essential spatial regularizer in the phase-field evolution law. Conversely, with increasing $\eta$ (e.g., $\eta = 0.2$ and $0.4$), the crack driving force stabilizes numerically as the hydrostatic stress and elastic strain fields contribute to the damage evolution. Physically, this transition represents a blended, energy-partitioned failure mechanism that couples both volumetric tensile dilatation and deviatoric plastic dissipation to represent realistic ductile degradation.

Conversely, for higher values ($\eta \ge 0.6$), macroscopic failure is inhibited within the investigated displacement range. Even though the individual magnitudes of $\tilde{\psi}^{\mathrm{e}+}$ and $\tilde{w}^\mathrm{p}$ may independently exceed the critical threshold $w_{\mathrm{c}}$, their weighted combination inside the Macaulay brackets in Eq.~(\ref{Eq: Ductile}) fails to overcome the fracture barrier. For instance, at $\eta = 0.6$, localized damage barely initiates with a peak phase-field value of only $\phi \approx 0.75$.

Fundamentally, the weighting factor $\eta$ acts as a regulatory parameter that partitions the relative contributions of elastic tensile energy and plastic dissipation to the total damage driving force. Because fracture processes are inherently driven by the local stress state and structural constraints, $\eta$ could vary between unnotched or smoothly notched tensile geometries—where unconstrained plastic flow dominates the failure mechanism —and highly constrained, sharp crack-tip configuration where failure is governed by intense hydrostatic stress concentrations.

\begin{figure}[hbt!]
  \includegraphics[width=12cm]{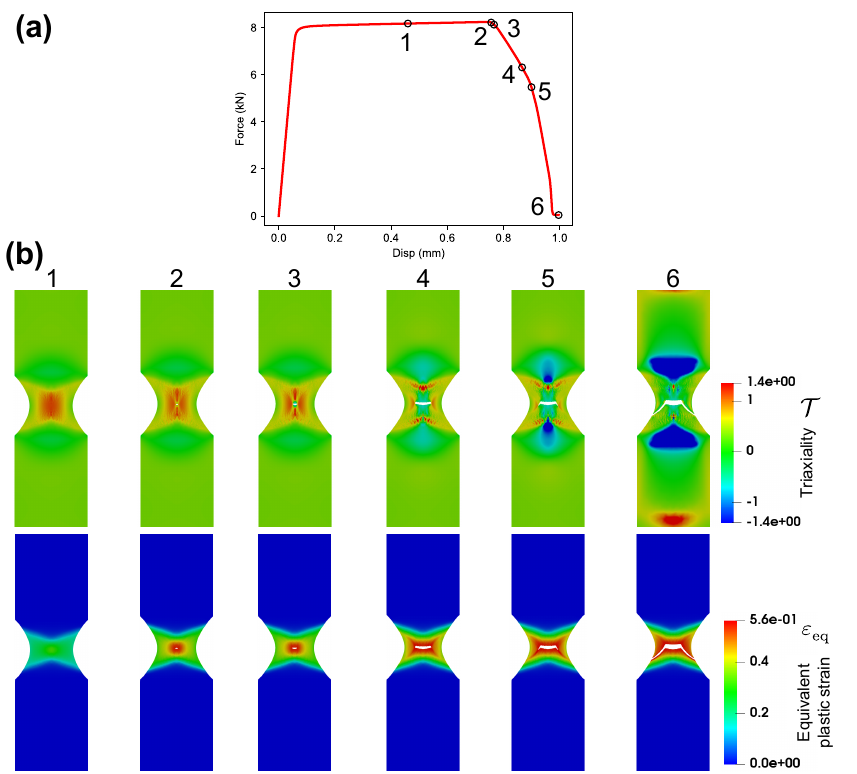}
  \centering
  \caption{ (a) load-displacement curve. (b) The triaxiality $\mathcal{T}$ and equivalent plastic strain $\varepsilon_\mathrm{eq}$ fields at different loading stages indicated in the force-displacement curve showing the crack evolution. Deformed configuration is scaled by a factor of 1.0 of the displacement vector field.}
  \label{Fig: NRB}
\end{figure}

Setting $\eta=0.4$, Fig.~(\ref{Fig: NRB}a) shows the resulting load-displacement curve, indicating the loading steps used for the contour plots of stress triaxiality $\mathcal{T}$ and equivalent plastic strain $\varepsilon_\mathrm{eq}$ fields presented in Fig.~(\ref{Fig: NRB}b). Within the notched region, the stress triaxiality is positive, reaching its peak at the specimen center, which is also the region of maximum plastic strain. Consequently, damage nucleates at this central location and propagates horizontally, perpendicular to the loading direction. Before the crack approaches the surface of the specimen in the remaining ligament, it deflects to $\approx 45^\circ$ angle forming a shear lip. This morphology successfully captures the characteristic 'cup-and-cone' fracture typical of ductile metals \cite{cao2014numerical}. 

\begin{figure}[hbt!]
  \includegraphics[width=12cm]{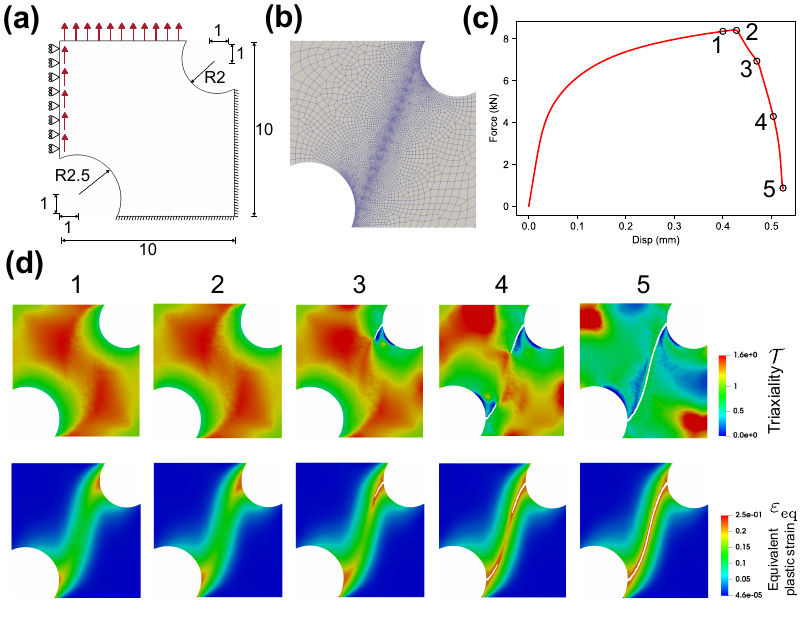}
  \centering
  \caption{(a) Geometry and boundary conditions of the double-notched specimen. (b) Finite element mesh (c) Force-displacement response. (d) The triaxiality $\mathcal{T}$ and equivalent plastic strain $\varepsilon_\mathrm{eq}$ fields at different loading stages indicated in the force-displacement curve showing the crack evolution. Deformed configuration is scaled by a factor of 1.0 of the displacement vector field.}
  \label{Fig: DoubleNotch}
\end{figure}

The second case considers the double-notched specimen proposed by Mediavilla et al. \cite{mediavilla2006robust}, which is widely employed as a benchmark for ductile damage models \cite{Ambati2015ductile, Alakheel2018}. The specimen geometry and applied boundary conditions are detailed in Fig.~(\ref{Fig: DoubleNotch}a) and the corresponding finite element mesh in Fig.~(\ref{Fig: DoubleNotch}b). For the material parameters, power-law hardening was used with $\sigma_{\mathrm{y}0} = 350$ MPa, $K=1300$ MPa and $n=0.1$. The phase-field fracture parameters were $\eta = 0.4$, $w_\mathrm{c} = 90 \, \mathrm{MJ/m^3}$.

A vertical displacement was applied to the left and top edges, while the right and bottom boundaries were held fixed. The resulting load-displacement response is shown in Fig.~(\ref{Fig: DoubleNotch}c), with the corresponding triaxiality and equivalent plastic strain maps in Fig.~(\ref{Fig: DoubleNotch}d) up to final damage. The failure sequence begins at the upper notch, where $\varepsilon_\mathrm{eq}$ is maximum and $\mathcal{T}$ is positive. This initial crack propagates in a curved downward path toward the center. Simultaneously, a second crack initiates at the lower notch and evolves upward. Final rupture is achieved when these two crack fronts coalesce at the specimen center, demonstrating excellent agreement with the benchmark results reported in \cite{Ambati2015ductile, Alakheel2018}.

\subsection{Tensile test with different hydrogen gas pressures}
\label{Section: Pressure}

In this section, we evaluate the effect of hydrogen gas pressure on the tensile behavior of a smooth round bar. The experimental data were obtained from Moro et al. \cite{Moro2010} for API X80 steel, which is widely used in oil and gas pipelines. Tensile tests were performed on smooth axisymmetric specimens with a diameter of 6 mm and gauge length of 30 mm. The specimens were subjected to nitrogen gas at a pressure of 30 MPa. The samples tested in hydrogen atmosphere were exposed to pressures of 0.1, 5, 10 and 30 MPa. All tests were performed at a strain rate of $5 \times 10^{-5} \, \mathrm{s}^{-1}$.

\begin{figure}[hbt!]
  \includegraphics[width=16cm]{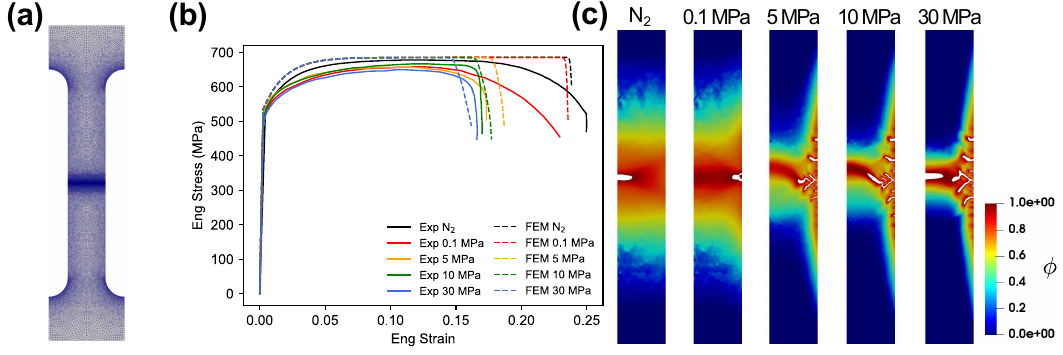}
  \centering
  \caption{The effect of surface hydrogen pressure on the tensile behavior of X80 steel: (a) Finite element mesh (b) The stress-strain curves compared to experimental data from Moro et al. \cite{Moro2010}. (c) The damage field at the point of failure showing the transition from localized damage at the center to surface cracking with increasing hydrogen pressure.}
  \label{Fig: Pressure}
\end{figure}

For the simulations, the hydrogen pressure was modeled as Dirichlet boundary condition with equilibrium concentration according to Eq.~(\ref{Eq: Conb}). $c_B$ was evaluated according to Sieverts' law $S_0\sqrt{P_H}$, where $S_0 = 6.2 \times 10^{-6} \, \mathrm{mol/(m^3 \sqrt{Pa})}$ \cite{Moro2010} and $P_H$ is the hydrogen gas pressure. A diffusivity value $D_\mathrm{L} = 2.059 \times 10^{-10} \, \mathrm{m^2/s}$ was used for API X80 following \cite{zheng2023effect}.
A Voce hardening model was used with initial yield strength $\sigma_{\mathrm{y}0} = 533$ MPa, saturation stress $\sigma_{\mathrm{sat}} = 685$ MPa, and initial strain hardening modulus $H_0 = 9$ GPa. $\eta = 0.4$ was used, while the hydrogen-dependent critical work density parameters in Eq.~(\ref{Eq: wc}) were $w_\mathrm{c0} = 90 \times 10^{6} \, \mathrm{J/m^3}$, $w_\mathrm{cmin} = 18 \times 10^{6} \, \mathrm{J/m^3}$, $c_\mathrm{crit} = 3.395 \, \mathrm{mol/m^3}$ and $\beta = 1$. The finite element mesh is shown in Fig.~(\ref{Fig: Pressure}a)

Fig.~(\ref{Fig: Pressure}b) shows the effect of hydrogen pressure on the engineering stress-strain curves derived from FEM simulations compared to the experimental results of Moro et al. \cite{Moro2010}. It can be observed that the ductility drops with increasing the hydrogen gas pressure and, consequently, surface concentration. The corresponding phase-field damage patterns $\phi$ are shown in Fig.~(\ref{Fig: Pressure}c), where values of $\phi > 0.98$ have been clipped to clearly visualize the crack paths. For the nitrogen environment reference case, the damage initiates at the center of the sample and propagates toward the surface, consistent with the general trends observed in the ductile failure of metals as discussed earlier.

Upon the introduction of 0.1 MPa pressure of hydrogen gas, the failure mode shifts; the damage initiates at the specimen surface and propagates toward the center. With further increasing the hydrogen pressure, the damage pattern transforms into multiple surface cracks within the central region of the sample. This behavior is in excellent agreement with the experimental trends reported by Moro el al. \cite{Moro2010} and several recent studies using in-situ hydrogen charging \cite{arniella2023hydrogen, malheiros2022local, zhang2024enhancement, santana2025investigating}.

Modeling strategies, such as the coupled Oriani-GTN frameworks developed by Pinto et al. \cite{Pinto2024simulation} and Fern{\'a}ndez-Pis{\'o}n et al. \cite{fernandez2026oxygen}, successfully captured surface cracking by implementing gradient regularization on the scalar accumulated plastic strain field, which directly governs local trap density evolution. In contrast, the present framework achieves this through an explicit transport-driven formulation (Eq.~(\ref{Eq: ConFlux})). By enforcing thermodynamic equilibrium concentration directly at the specimen boundary nodes (Eq.~(\ref{Eq: Conb})), a highly localized surface saturation zone is established. As plastic deformation progresses near the surface, a steep spatial gradient of the normalized dislocation density ($\nabla\bar{\rho}$) develops. Because the local defect density is a direct function of equivalent plastic strain (Eq.~(\ref{Eq: KMEAnalytical})), this spatial gradient term is implicitly proportional to the plastic strain gradient.

\begin{figure}[hbt!]
  \includegraphics[width=16cm]{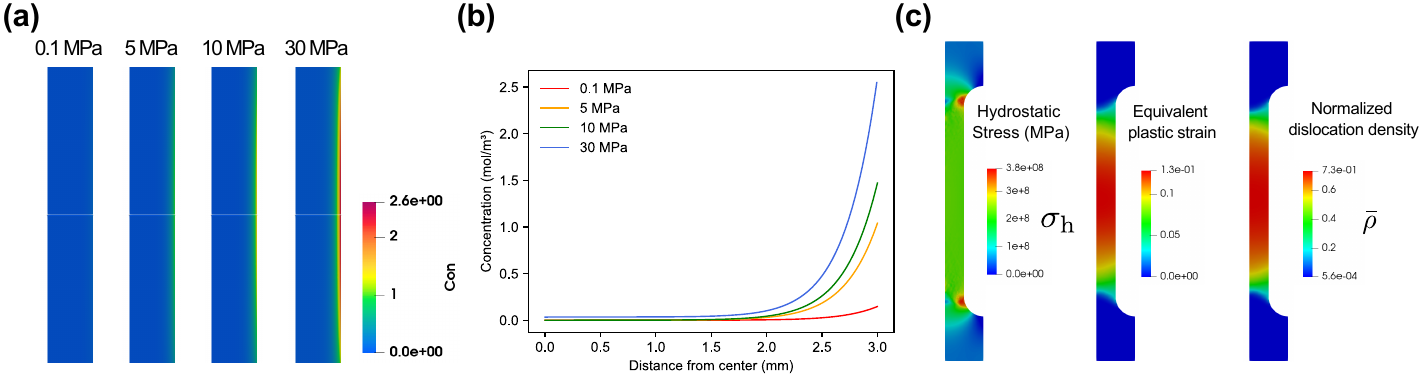}
  \centering
  \caption{(a) Spatial distribution of hydrogen concentration at an applied strain of $\varepsilon = 0.102$ for various hydrogen pressures. (b) Radial concentration profiles extracted at the specimen center. (c) The corresponding contour plots of hydrostatic stress $\sigma_\mathrm{h}$, equivalent plastic strain $\varepsilon_\mathrm{eq}$ and normalized dislocation density $\bar{\rho}$. Note that these mechanical fields are representative for all concentration cases at this loading stage, as damage initiation has not yet occurred.}
  \label{Fig: PressureCon}
\end{figure}

In order to understand the shift in damage initiation from the core to the surface, we investigate the hydrogen distribution. Fig.~(\ref{Fig: PressureCon}a) illustrates the hydrogen concentration at an applied strain of $\varepsilon = 0.102$. At this loading stage, damage has not yet initiated; consequently, the stress, strain, and all derived mechanical quantities remain identical across all test cases. The corresponding concentration profiles at the specimen mid-section are shown in Fig.~(\ref{Fig: PressureCon}b). As expected, hydrogen concentration increases from $\approx$ zero at the specimen axis of symmetry to the equilibrium value $c_\mathrm{eq}$ prescribed by Eq.~(\ref{Eq: Conb}) at the surface. This results in a \emph{skin effect}, characterized by a region of high hydrogen concentration near the surface that intensifies with increasing the surface concentration.

However, these profiles are not uniform along the gauge length, as the local hydrogen accumulation follows the gradients of the hydrostatic stress $\sigma_\mathrm{h}$ and the normalized dislocation density $\bar{\rho}$. Contour plots of these fields are provided in Fig.~(\ref{Fig: PressureCon}c). While the hydrostatic stress remains relatively uniform throughout the gauge length with maximum values at the upper and lower fillets, the equivalent plastic strain—and by extension, the normalized dislocation density—gradually localize toward the specimen center reaching a maximum of 0.73, which corresponds to an absolute dislocation density of $\approx 1.3 \times 10^{14}\text{ m}^{-2}$. In the hydrogen-free case, this localization of plastic work density $\tilde{w}^\mathrm{p}$ in the center at the core triggers damage initiation from the center outward.

In the presence of hydrogen, this localization makes the skin effect non-uniform along the gauge length, reaching a maximum at the specimen's mid-section. This is clearly visible in the 30 MPa concentration profile in Fig.~(\ref{Fig: PressureCon}a). These results highlight that accounting for hydrogen segregation at dislocations is crucial for predicting hydrogen-mediated damage patterns; relying solely on hydrostatic stress-driven diffusion is insufficient to reproduce the observed failure morphology. 

\begin{figure}[hbt!]
  \includegraphics[width=16cm]{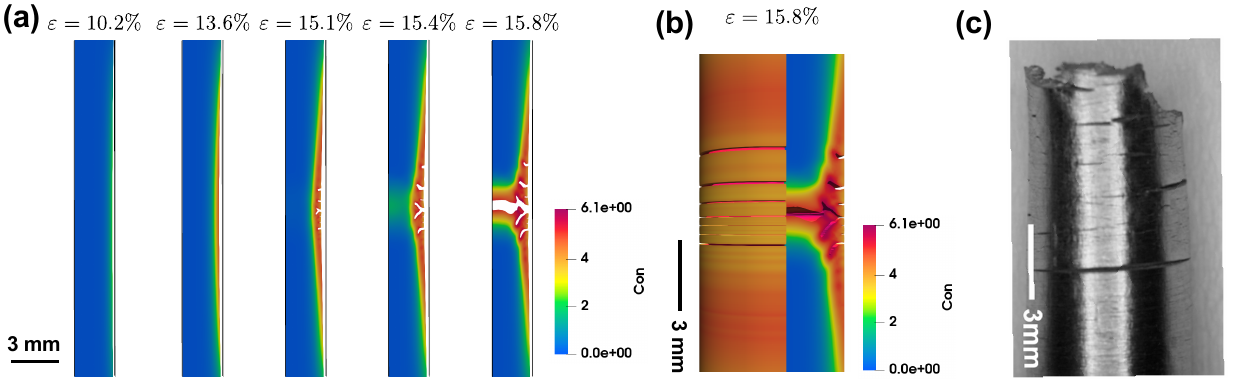}
  \centering
  \caption{(a) Simulated evolution of the hydrogen concentration field at several loading stages, showing the development of surface cracks for the 30 MPa case. Deformed configuration is scaled by a factor of 1.0 of the displacement vector field. (b) A 3D rotational extrusion of the specimen at $\varepsilon = 15.8 \%$, illustrating the circumferential nature of the surface cracking (c) Comparison with experimental observations showing morphological agreement with the predicted surface cracking patterns in the necked region of an API X80 steel specimen, adapted from Moro et al. \cite{Moro2010} with permission from Elsevier.}
  \label{Fig: Evolve30MPa}
\end{figure}

To understand the formation mechanism of this hydrogen-Induced multiple surface cracking, the evolution at several loading stages for the 30 MPa case is shown in Fig.~(\ref{Fig: Evolve30MPa}a). The contours are plotted in the deformed configuration, with solid black lines representing the undeformed geometry for reference. A 3D rotational extrusion view at $\varepsilon = 15.8 \%$ is shown in Fig.~(\ref{Fig: Evolve30MPa}b), showing a remarkable morphological resemblance to the experimental observations of circumferential cracking rings reported by Moro et al. \cite{Moro2010} (Fig.~\ref{Fig: Evolve30MPa}c).

From Fig.~(\ref{Fig: Evolve30MPa}a), it can be seen that these surface cracks develop mainly in the central region of the gauge length, where plastic deformation is maximum. This phenomenon arises from a competition between the relatively ductile interior of the specimen and the embrittled surface skin effect discussed earlier. Such failure mode arises from a mechanical incompatibility between the ductile interior and the embrittled near-surface skin region, where the hydrogen-induced reduction of $w_{\mathrm{c}}$ drives the initiation of multiple circumferential surface cracks. This mechanism is similar to thermal shock cracks in ceramics \cite{bourdin2014morphogenesis, SICSIC2014256, wang2024phase, li2022three}.

\subsection{Effect of strain rate on tensile test with in-situ hydrogen charging}

\begin{figure}[hbt!]
  \includegraphics[width=15cm]{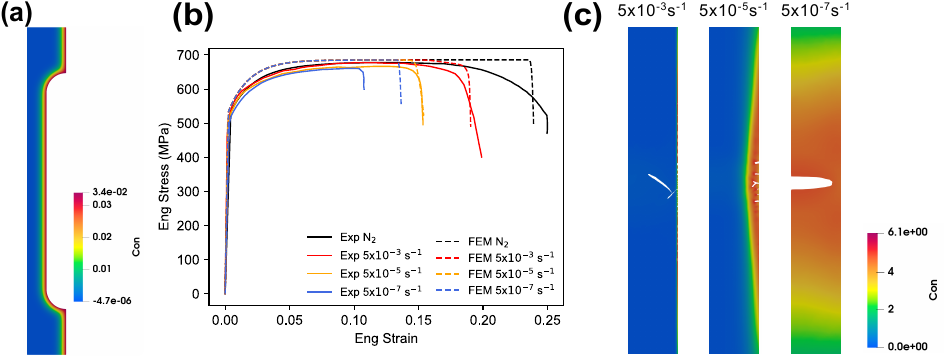}
  \centering
  \caption{(a) Hydrogen concentration after pre-charging a stress-free specimen for 30 min at 30 MPa.} (b) Effect of strain-rate on the tensile behavior at constant hydrogen gas pressure of 30 MPa compared to experimental data of API X80 steel from Moro et al. \cite{Moro2010}. (c) Contours of hydrogen concentration and corresponding damage patterns at failure for different loading rates.
  \label{Fig: Rate}
\end{figure}

In this section, we investigate the influence of the applied strain rate on hydrogen embrittlement. The experimental results for comparison are also obtained from Moro et al. \cite{Moro2010}. All specimens were pre-charged under a hydrogen gas pressure of 30~MPa for 30~min \cite{Moro2010}. Numerically, this pre-charging step was modeled using a stress-free geometry for a duration of 30~min, where a constant boundary concentration $c_{\mathrm{B}}$ is prescribed at the outer surface according to Sieverts' law. Fig.~(\ref{Fig: Rate}a) shows the concentration profile at the end of the pre-charging simulations, which is the initial state for the loading simulations. Subsequently, tensile tests were performed in-situ under the same hydrogen environment at strain rates of $5 \times 10 ^{-3} \, \mathrm{s}^{-1}$, $5 \times 10 ^{-5} \, \mathrm{s}^{-1}$ and $5 \times 10 ^{-7} \, \mathrm{s}^{-1}$. The material properties are similar to those specified in Section \ref{Section: Pressure}.

The stress-strain results presented in Fig.~(\ref{Fig: Rate}b) show a progressive loss of ductility with decreasing strain rate. Fig.~(\ref{Fig: Rate}c) provides the corresponding contour plots of hydrogen concentration and damage patterns. At the highest strain rate ($5 \times 10 ^{-3} \, \mathrm{s}^{-1}$), the damage initiates at the surface, showing several shallow circumferential cracks. These cracks remain localized near the skin due to the limited time available for hydrogen diffusion during the rapid loading. This eventually leads to the formation of a main crack resulting in failure.

The intermediate case ($5 \times 10 ^{-5} \, \mathrm{s}^{-1}$) exhibits a damage pattern similar to that discussed in Section~\ref{Section: Pressure}; however, the failure strain drops from $\approx$ 16 \% to 15 \% due to the additional hydrogen uptake during the 30-minute pre-charging period. Finally, the slowest rate ($5 \times 10 ^{-7} \, \mathrm{s}^{-1}$)  results in a single crack propagating from the center of the specimen toward the surface, a morphology similar to the hydrogen-free reference case Fig.~(\ref{Fig: Pressure}b). This occurs because the extremely low loading rate allows sufficient time for hydrogen to diffuse uniformly throughout the specimen cross-section. Consequently, the critical work density $w_\mathrm{c}$ becomes spatially uniform and reaches its degraded state according to Eq.~(\ref{Eq: wc}), eliminating the concentration gradients that drive surface initiation in the faster cases. Similar mechanism has been proposed using a GTN model by Pinto et al. \cite{Pinto2024simulation} and Fern{\'a}ndez-Pis{\'o}n et al. \cite{fernandez2026oxygen}. These results clearly illustrate the transition from surface-limited cracking at high strain rates to bulk-dominated fracture at lower rates.

\subsection{Fracture toughness test in hydrogen atmosphere}

We demonstrate the application of the modeling framework to fracture toughness test of compact tension (CT) specimens for carbon steel. The experimental data for these tests were reported by Ogawa et al. \cite{ogawa2017unified}. Following \cite{san2012technical}, a Sieverts' law constant of $S_0 = 6.2 \times 10^{-6} \, \mathrm{mol/(m^3 \sqrt{Pa})}$ was used, along with a lattice diffusivity of $D_\mathrm{L} = 1.137 \times 10^{-8} \, \mathrm{m^2/s}$. A power-law hardening was used with $\sigma_{\mathrm{y}0} = 360$, $K=550$ MPa and $n=0.1$. The phase-field fracture parameters were set to $\eta = 0.6$, $w_\mathrm{c0} = 60 \times 10^{6} \, \mathrm{J/m^3}$, $w_\mathrm{cmin} = 9 \times 10^{6} \, \mathrm{J/m^3}$, $c_\mathrm{crit} = 1.08 \, \mathrm{mol/m^3}$ and $\beta = 2$.

\begin{figure}[hbt!]
  \includegraphics[width=10cm]{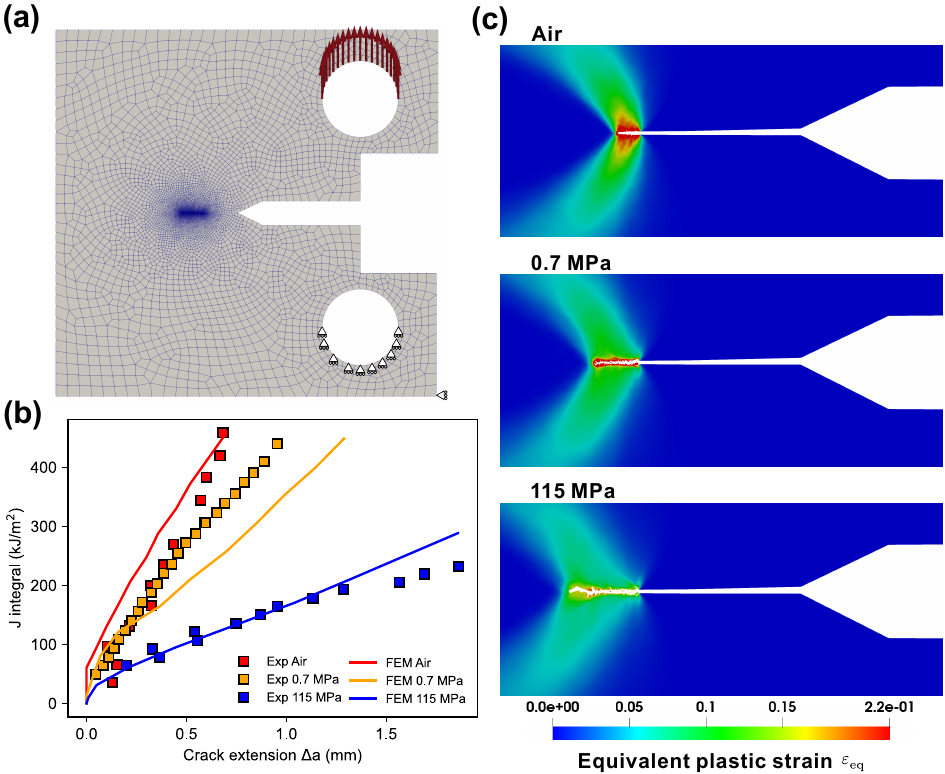}
  \centering
  \caption{(a) Compact tension specimen showing the boundary conditions and finite element mesh. (b) J-resistance curves plots for different in-situ hydrogen exposures compared to the experimental results of Ogawa et al. \cite{ogawa2017unified} for carbon steel. (c) Full field results of the equivalent plastic strain $\varepsilon_\mathrm{eq}$ illustrating the crack extension and the evolution of the fracture process zone at the end of loading. Deformed configuration is scaled by a factor of 0.2 of the displacement vector field.}
  \label{Fig: CT}
\end{figure}

Fig.~(\ref{Fig: CT}a) shows the Compact Tension (CT) specimen, with the applied boundary conditions and finite element mesh. All simulations were performed at a constant displacement rate of $2 \times 10^{-3}$ mm/s. The crack extension was measured during post-processing. The resulting J-resistance curves for the three exposure cases—Air, 0.7 MPa and 115 MPa of hydrogen gas—are presented in Fig.~(\ref{Fig: CT}b). The numerical predictions show excellent agreement with the experimental results of Ogawa et al. \cite{ogawa2017unified}, where the fracture resistance decreases with increasing hydrogen pressure.

This degradation is further illustrated by the equivalent plastic strain $\varepsilon_\mathrm{eq}$ contours and the corresponding crack extension at the final loading stage in Fig.~(\ref{Fig: CT}c). In the absence of hydrogen, a large, well-developed plastic zone is observed ahead of the crack tip, characteristic of high-toughness ductile tearing. Conversely, as hydrogen pressure increases, the plastic strain becomes highly localized along the crack propagation path, signaling a transition toward an embrittled fracture mode. This shift is quantitatively reflected in the crack extension $\Delta \mathrm{a}$, which increases nearly fourfold at 115 MPa compared to the Air case for the same loading level. Furthermore, the simulated crack paths in the hydrogen-charged specimens exhibit a more irregular, 'jagged' morphology. This is qualitatively consistent with experimental observations \cite{olden2009influence, lee2025direct}, which highlight how hydrogen-induced slip localization can deviate the crack path from the ideal opening plane.

\FloatBarrier

\section{Conclusions}

In this work, a comprehensive chemo-mechanical framework was developed by coupling a fully kinetic hydrogen transport model with a modular geometric phase-field fracture method to investigate hydrogen embrittlement (HE) in metals. The formulation was specifically designed to resolve the interplay between hydrogen-dislocation interactions and stress-state-dependent ductile failure, which were shown crucial to represent characteristic morphologies of HE. The simulation results showed excellent agreement with experimental results in the literature, especially in smooth tensile specimens. The primary findings and contributions of this study are summarized as follows:

\begin{itemize}
  \item A phenomenological crack driving force was proposed based on the weighted contributions of the positive (tensile) part of the elastic strain energy and plastic work densities. By employing a hyperbolic tangent function of the stress triaxiality, the model ensures that plastic dissipation contributes to the fracture process only in tensile regions, effectively mimicking the physics of void-driven ductile damage. This formulation successfully captured characteristic failure morphologies, such as the cup-and-cone transition in notched specimens, while offering computational efficiency and ease of calibration compared to traditional micromechanical models like GTN.
  \item For smooth round bars under in-situ charging, the model predicted a shift in damage initiation as a function of hydrogen pressure. While hydrogen-free samples exhibited center-initiated failure, the introduction of hydrogen triggered surface initiation that evolved toward the core. At higher pressures, a "skin effect" emerged, resulting in multiple circumferential surface cracks around the central necking region of the specimens. This phenomenon is driven by the mechanical incompatibility between a ductile specimen core and a hydrogen-embrittled surface layer, a behavior that could only be resolved through the inclusion of hydrogen segregation at dislocations.
  \item The framework demonstrated a strong sensitivity to the applied loading rate, capturing a transition from multiple surface cracking at higher rates to a single, center-initiated flat crack at extremely low rates. This transition highlights a kinetic competition: at low loading rates, sufficient time is available for hydrogen to diffuse uniformly across the cross-section, eliminating the steep concentration gradients that drive surface cracking at faster rates.
  \item The model was further validated against experimental J-resistance curves for CT specimens. The results demonstrate that the localized reduction in critical work density accurately characterizes the loss of fracture toughness and the shift from widespread ductile tearing to localized embrittled crack propagation.
\end{itemize}

\section*{Acknowledgements}

This work is supported by the University of Oulu and the Research Council of Finland Profi 352788 through their funding of the H2Future project. VJ and JK would like to thank Jane and Aatos Erkko (J\&AE) Foundation and Tiina and Antti Herlin (TAH) Foundation for their financial supports of Advanced Steels for Green Planet project (AS4G).

\section*{Declaration of generative AI and AI-assisted technologies in the manuscript preparation process}

During the preparation of this work the author(s) used Gemini in order to refine the technical language and improve the grammatical clarity of the manuscript. After using this tool/service, the author(s) reviewed and edited the content as needed and take(s) full responsibility for the content of the published article.

\bibliographystyle{unsrtnat}
\bibliography{references}

\end{document}